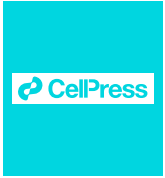
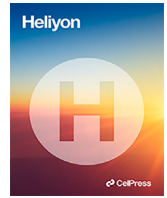

Research article

# Combination of frequency- and time-domain characteristics of the fibrillatory waves for enhanced prediction of persistent atrial fibrillation recurrence after catheter ablation

Pilar Escribano [a],[*], Juan Ródenas [a], Manuel García [a], Miguel A. Arias [b], Víctor M. Hidalgo [c], Sofía Calero [c], José J. Rieta [d], Raúl Alcaraz [a]

[a] *Research Group in Electronic, Biomedical and Telecommunication Engineering, University of Castilla-La Mancha, Albacete, Spain*
[b] *Cardiac Arrhythmia Department, Complejo Hospitalario Universitario de Toledo, Toledo, Spain*
[c] *Cardiac Arrhythmia Department, Complejo Hospitalario Universitario de Albacete, Albacete, Spain*
[d] *BioMIT.org, Electronic Engineering Department, Universitat Politecnica de Valencia, Valencia, Spain*

A B S T R A C T

Catheter ablation (CA) remains the cornerstone alternative to cardioversion for sinus rhythm (SR) restoration in patients with atrial fibrillation (AF). Unfortunately, despite the last methodological and technological advances, this procedure is not consistently effective in treating persistent AF. Beyond introducing new indices to characterize the fibrillatory waves ($f$-waves) recorded through the preoperative electrocardiogram (ECG), the aim of this study is to combine frequency- and time-domain features to improve CA outcome prediction and optimize patient selection for the procedure, given the absence of any study that jointly analyzes information from both domains. Precisely, the $f$-waves of 151 persistent AF patients undergoing their first CA procedure were extracted from standard V1 lead. Novel spectral and amplitude features were derived from these waves and combined through a machine learning algorithm to anticipate the intervention mid-term outcome. The power rate index ($\varphi$), which estimates the power of the harmonic content regarding the dominant frequency (DF), yielded the maximum individual discriminant ability of 64% to discern between individuals who experienced a recurrence of AF and those who sustained SR after a 9-month follow-up period. The predictive accuracy was improved up to 78.5% when this parameter $\varphi$ was merged with the amplitude spectrum area in the DF bandwidth ($AMSA_{LF}$) and the normalized amplitude of the $f$-waves into a prediction model based on an ensemble classifier, built by random undersampling boosting of decision trees. This outcome suggests that the synthesis of both spectral and temporal features of the $f$-waves before CA might enrich the prognostic knowledge of this therapy for persistent AF patients.

## 1. Introduction

Atrial fibrillation (AF) is a heart condition that overrides normal sinus rhythm (SR) due to uncoordinated atrial electrical activation causing ineffective contractions of the atria [1,2]. In clinical practice, this is the most frequently encountered cardiac arrhythmia [3], showing an growing incidence and prevalence with advancing age [4]. In fact, the estimated prevalence in adults ranges from 2 to 4%, and it is anticipated to rise 2.3-fold due to the aging population and search intensification for undiagnosed individuals [2]. This situation places AF as one of the most significant epidemics and challenges in public health [5].

Normal cycle depolarization of cardiac cells during SR proceeds in an orderly fashion beginning in the sinus node and then spreading sequentially through the atria, atrioventricular node, and ventricles. This sequential process results in the consecutive P-wave, QRS complex, and T-wave on the electrocardiogram (ECG) signal [3]. Contrarily, AF involves abnormal generation and transmission of wavefronts through the atrial substrate, which are recorded in the ECG as an undulatory, inconsistent activity that replace the P-wave by many fibrillatory waves ($f$-waves). However, the QRS complex is still present in the ECG, presenting irregular occurrence over time due to the chaotic propagation of the atrial wavefronts to the ventricles [3]. Some common symptoms associated with this anomalous behavior of the heart are chest pain, fatigue, dizziness, inappropriate heart rate acceleration, shortness of breath, and palpitation attacks, among others [6]. Although AF itself is not recognized as a primary cause of death, patients face an elevated risk of ischemic stroke and heart failure that rises two-fold the mortality [7]. This situation is more dangerous for those patients with obesity, obstructive sleep apnea, hypertension, diabetes mellitus, and alcohol consumption, conditions that have also been identified in themselves as risk factors of AF [6,8].

The still incomplete knowledge of AF aetiology makes selection of the most suitable therapy for each arrhythmia stage a challenge for clinicians [1]. However, AF management guidelines recommend different alternatives depending on current AF classification, which is based on how the arrhythmia manifests [2]. Thus, paroxysmal AF is usually the earliest phase, in which arrhythmic episodes end spontaneously within a week. It is followed by a persistent stage, when the arrhythmia fails to self-terminate requiring clinical intervention to restore SR. Finally, permanent AF is positioned at the end and most advanced stage. In this case, the physician and patient jointly agree to stop additional attempts to restore SR because of the strong permanence of the disease [2]. Nevertheless, each patient frequently presents different symptoms and treatment outcomes regardless of their AF classification [6]. Besides, AF frequently evolves through its different stages and is associated with a remodeling process affecting the structure and electrical properties of the atrial substrate that promotes its perpetuation [9]. Hence, early diagnosis and restoration of SR is totally recommended to avoid the progression of the disease to more advanced stages [10].

For that purpose, catheter ablation (CA) stands as the primary treatment for patients experiencing symptomatic, drug-refractory AF [11], because it is associated with higher efficacy than antiarrhythmic therapy for maintaining SR [12]. This intervention is a minimally invasive procedure mostly focused on the pulmonary vein isolation (PVI) technique to stop the abnormal electrical conduction in the atrial tissue [13]. It pursues a permanent therapeutic solution to enhance the patient's quality of life and, in most cases, avoid a chronic antiarrhythmic treatment [3]. Although this outcome is achieved for many patients in early stages of AF, between one-third and one-half of those in a persistent stage relapsed to AF within one year after the first PVI [14,15], then requiring additional CA procedures that in many cases are also ineffective [1].

The increment of hospitalizations and emergency out-patient visits associated with AF patients [8] and the complexity of the CA procedure, which entails time-consuming interventions [16], explain the significant burden that AF places on health systems worldwide [2]. This situation has prompted a clinical focus on predicting CA outcome before the procedure, with the aim of selecting individuals who truly experience benefits from the treatment and reduce unnecessary risks and assign more appropriate AF treatments as soon as possible to the rest [17]. Hence, such prediction could reduce hospitalizations, limit repeated CA procedures, improve quality of life of AF patients, and alleviate the burden and costs associated with the arrhythmia treatment, among others [16,17].

Different clinical indices mostly obtained from the patient's lifestyle, demographic characteristics, medical history, and anatomical and functional heart parameters have been proposed as post-CA AF recurrence risk factors. However, they have provided controversy results as well as limited predictive ability [18]. The combination of some of these variables through traditional statistical approaches (i.e., mainly logistic regression) has also been explored to build scoring systems. Again, they have reported largely suboptimal results, with limited generalization ability on unknown populations [15]. In line with the current trend that modern machine learning (ML) algorithms are able to provide more general and predictive cardiovascular models of clinical variables than traditional scores [19–21], a recent ML-based method has shown an improved risk prediction of one-year AF recurrence after CA regarding currently available tools [22]. However, this model loses clinical interpretability of its predictions, since it is based on combining nineteen variables. As well, its predictive ability on unknown data is still far from being clinically optimal, with a global performance of about 72%.

In the last years, there is a growing focus on analyzing $f$-waves extracted from the ECG signal to preoperatively anticipate CA outcome [23]. Well-known time and frequency parameters of these waves, including their amplitude and dominant frequency (DF), are today considered as surrogate markers of remodeling in the atrial substrate and have proven to be effective in predicting the success of different AF treatments [24,25]. In the particular case of CA, they have reported a higher performance than clinical parameters and scores, but there is still room for improvement [23]. Moreover, predictive models based only on traditional statistical approaches have been mainly studied to combine detachedly frequency [26] or time [27,28] features, largely leaving unexplored the possible complementary information between both domains of the $f$-waves. Hence, the present work aims at broadening the time and frequency characterization of the $f$-waves to obtain novel features and exhaustively explore their combination with modern ML-based techniques to enhance CA mid-term result prediction from patients with persistent AF.

## 2. Materials and methods

This section introduces the population recruited for analysis and the CA protocol and follow-up strategy applied to all the patients. Then, the algorithms used for ECG preprocessing, as well as for extraction and characterization of the $f$-waves are presented. Finally, statistical analysis of the results and performance assessment of the proposed ML-based predictive models for the prediction of CA outcome are explained. As a summary, Fig. 1 shows a global overview of the study methodology. All details are provided in the next subsections.

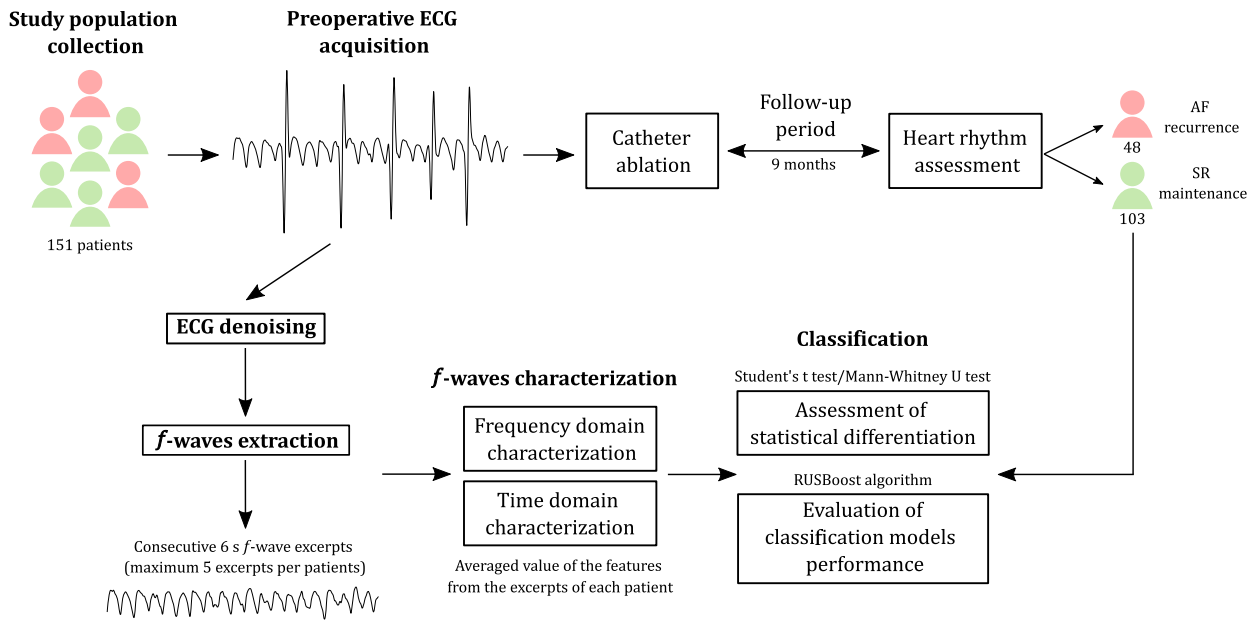

**Fig. 1.** Methodology diagram of the study.

*2.1. Study population*

This work enrolled 151 patients (35 women and 116 men) diagnosed with persistent AF, with an average age of 59 years old, raging from 20 to 82. They were selected under standard clinical indications to undergo their first radiofrequency CA treatment at the University Hospitals of Albacete and Toledo in Spain. Since this study involves human subjects, it complies with the ethical principles defined by the Declaration of Helsinki.

*2.2. Ablation protocol*

Antiarrhythmic drug therapy, excepting amiodarone, was discontinued in all patients more than 5 days (i.e., more than 5 half-lives) before the intervention. The patients remained under sedation during the procedure, either through the use of conscious sedation or general anesthesia. Moreover, anticoagulant drugs were supplied during the procedure to prevent thromboembolic complications. This involved an initial heparin bolus, with subsequent doses adjusted based on activated coagulation time monitoring. Catheters were introduced through the femoral venous access and transseptal puncture was carried out to achieve the left atrium, where the location of PVs was determined using a mapping catheter. Later, PVI technique was done by placing a radiofrequency catheter in the atrial tissue delivering current for at least 30 seconds [29]. The main objective of that procedure consisted of the generation of point by point lesions, encircling the PVs and creating an electrically impenetrable boundary to isolate the arrhythmogenic region from the rest of the heart substrate [30]. Finally, the procedure reached its conclusion with the successful isolation of all PVs or, if in that point AF still remained, after restoring SR by electrical cardioversion.

*2.3. Follow-up and procedure outcome*

The intervention was successfully completed in all patients, who did not experience relevant complications during a monitoring period of several hours after the ablation. Along the 9-month follow-up period, the patients were administered anticoagulant and antiarrhythmic drugs at the discretion of the healthcare professionals and maintained the recommended visit and monitoring protocol outlined in the current AF management guidelines [2]. At the end of this period, out of the 151 patients included in the study, only 103 maintained SR, whereas 48 experienced a relapse into AF. Hence, according to current recurrence statistics [13,14], CA was unsuccessful for approximately 32% of the enrolled persistent AF patients. From now on, a differentiation will be made between both groups of patients, i.e., between individuals who relapsed to AF and those who maintained SR after the follow-up period.

*2.4. Preprocessing of the ECG signal*

Cardiac activity of the patient was continuously monitored during the entire CA intervention, acquiring a common 12-lead ECG signal, sampled at 977 Hz and 16 bits. In particular, the preoperative ECG interval was selected from the recording of the lead

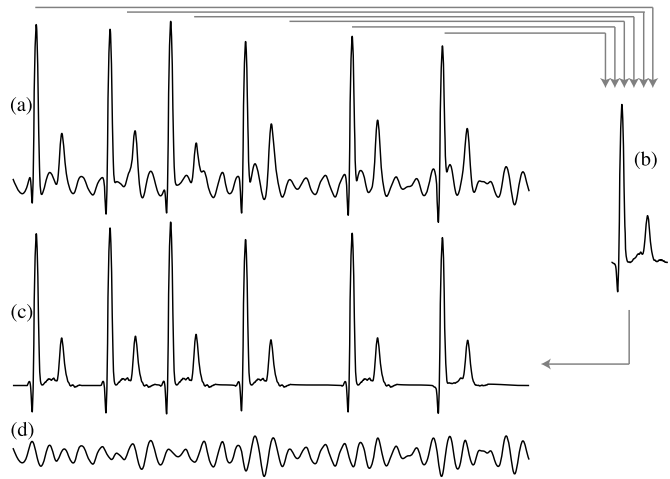

**Fig. 2.** Extraction of *f*-waves from an ECG in AF using a variant of the average beat subtraction method [35]. (a) The preprocessed ECG signal. (b) QRST complex template for cancellation. (c) Ventricular cancellation template. (d) The extracted *f*-waves.

V1 for each patient. This lead was chosen for the subsequent analysis of the fibrillatory activity due to its nearness to the right atrium. This favors the capture of the greatest amplitude *f*-waves in relation to ventricular activity, then facilitating their further extraction [31]. Before this step, the interval was preprocessed to remove the baseline wander, the 50 Hz powerline interference, and other possible high frequency disturbances [32]. Precisely, the ECG low frequency wandering was estimated with a 0.8 Hz cut-off frequency low-pass bidirectional filtering strategy and subsequently extracted from the initial signal [32]. Next, an stationary wavelet transform-based algorithm was applied to eliminate the powerline interference [33]. Although this method can substantially reduce the high-frequency noise, a 70 Hz cut-off frequency low-pass bidirectional filter was also considered to get the cleanest possible signal [32].

The number of *f*-waves seen in the TQ interval, where the ventricles do not interfere, varies with the heart rate and how often the *f*-waves repeat [34]. However, a more accurate characterization has been obtained when these waves are extracted from the preprocessed ECG recording [34]. Although several methods exist for that purpose, one of the most known is the average beat subtraction [34]. The variant named adaptive singular value cancellation of the QRST complexes was used in the present work. As Fig. 2 shows, this algorithm initially performed the singular value decomposition on a collection of temporally aligned QRST segments of the signal, considering the previously detected R peaks as reference points, to obtain the cancellation template. For all the beats, seven or more similar complexes were available, thus ensuring successful performance [35]. Later, the template was adapted for each QRST complex cancellation in amplitude and the discontinuities between complexes and the original signal originated from the cancellation were softened by Gaussian fitting [35]. Visual supervision finally corroborated absence of high ventricular residues in extracted *f*-waves.

### 2.5. Characterization of the *f*-waves

Once *f*-waves were extracted, the resulting signal from every patient was divided into non-overlapped excerpts of 6 s-length. A limit of 5 consecutive excerpts per patient was established due to the unequal duration of the acquired ECG intervals. Then, to fully describe fibrillatory activity, several frequency- and time-domain parameters were computed for each segment. Finally, the values obtained were averaged to obtain more robust results. In this way, it was performed an analysis and classification based on the subject and independent of the duration of the ECG signal. Moreover, moderate intra-patient variations observed in some of the analyzed parameters for periods as short as a few seconds [24,36] were minimized. It should be noted that the segment size of 6 s was empirically selected, after conducting several experimental tests considering recommendations from previous works [37–39]. No great differences were noticed in classification outcomes for segments of between 6 and 15 s in length, and the first value was selected to maximize the set of patients to be analyzed. Some previous works have also analyzed time and frequency features *f*-waves from similar or shorter ECG portions, e.g. [40–42].

#### 2.5.1. Spectral parameters computed from the *f*-waves

Spectral analysis of *f*-waves has been the focal point of several recent studies dealing with preoperative CA success prediction [23]. Hence, in the present work the extracted 6 s-length intervals of *f*-waves were characterized in the frequency-domain through common features, but also using novel parameters that have been introduced here for the first time. All these features were automatically obtained from the power spectral density (PSD) of the *f*-waves segments, hereinafter referred as $\mathcal{W}(f)$. The estimation of $\mathcal{W}(f)$ was computed through the Welch Periodogram, using a 4,000 points in length Hamming window with an overlap of 3,000 points between adjacent windowed sections, and 0.1 Hz spectral resolution.

A common reference metric that has already reported promising predictive results is the DF, which was named $f_0$ and determined as the frequency exhibiting a maximum in PSD amplitude [24,43,44]. To minimize the impact of noise and other artifacts, it was corroborated that $f_0$ was always estimated within the most common range for persistent AF, i.e., between 5 and 12 Hz [45]. The first harmonic of the DF, $f_1$, has also been analyzed in previous studies pursuing the same goal and was measured using a 1 Hz bandwidth centered on $2 \cdot f_0$. This metric was included in this analysis together with the spectral power of both, $f_0$ ($\mathcal{W}(f_0)$) and $f_1$ ($\mathcal{W}(f_1)$). In the same line, the DF bandwidth ($b_0$), limited between those lower and upper frequencies close to $f_0$ where the spectral power is halved from $\mathcal{W}(f_0)$, and the spectral power content in this bandwidth ($\mathcal{W}(b_0)$) were also considered [26]. Additionally, the harmonic exponential decay ($\gamma$) of the $f$-waves, that is given by the equation (1), was also included in the study as a reference parameter, since it achieved encouraging outcomes in earlier research [26]. It estimates the relevance of the DF harmonic content and was computed as the logarithmic ratio of $f_0$ and $f_1$ spectral power, i.e.,

$$\gamma = \ln\left(\frac{\mathcal{W}(f_0)}{\mathcal{W}(f_1)}\right) \tag{1}$$

Another spectral feature previously analyzed by Alcaraz et al. [26] was the median frequency ($f_m$), which represents the frequency that divides the whole PSD in two equal regions, as described in equation (2):

$$\sum_{f=3\text{ Hz}}^{\mathcal{M}} \mathcal{W}(f) = \sum_{f=\mathcal{M}}^{25\text{ Hz}} \mathcal{W}(f) = \frac{1}{2} \sum_{f=3\text{ Hz}}^{25\text{ Hz}} \mathcal{W}(f) \tag{2}$$

Besides, the centroid frequency ($f_c$) and the centroid power ($\mathcal{W}(f_c)$), given by equations (3) and (4), respectively, represent where is located the center of mass of a distribution, whose values are the PSD frequencies for $f_c$ and the PSD amplitudes for $\mathcal{W}(f_c)$, and the probabilities to observe these are the normalized PSD amplitude. These last two parameters were targeted in other works as defibrillation outcome predictors [46,47] and, for the first time, they were calculated for predicting CA outcome in the present study.

$$f_c = \frac{\sum_{f=3\text{ Hz}}^{25\text{ Hz}} \mathcal{W}(f) f}{\sum_{f=3\text{ Hz}}^{25\text{ Hz}} \mathcal{W}(f)} \tag{3}$$

$$\mathcal{W}(f_c) = \frac{\sum_{f=3\text{ Hz}}^{25\text{ Hz}} \mathcal{W}^2(f)}{\sum_{f=3\text{ Hz}}^{25\text{ Hz}} \mathcal{W}(f)} \tag{4}$$

These $f$-wave spectral characteristics were computed in the bandwidth between 3 and 25 Hz, where typical spectral content of the $f$-waves is concentrated [26]. Nevertheless, other spectral metrics computed in this work have estimated the $f$-wave organization considering three different frequency bands. On the one hand, the bandwidth between 3 and 25 Hz was divided into two consecutive frequency bands, encompassing the spectral content around the DF of the $f$-waves and its harmonic content, namely low-frequency (LF) and high-frequency (HF) bands, respectively. To clearly divide them, the position of the cut-off frequency was established in the midpoint between the DF and its first harmonic, and was computed as three means of the $f_0$ value from the origin of the spectrum. On the other hand, a total-frequency (TF) band comprising the spectral content of both LF and HF bands (3–25 Hz) was also analyzed to provide a global overview of the $f$-waves spectral distribution.

Additionally, several spectral power ratios were also included to evaluate the spectral organization, such as the organization index ($\mathcal{O}$) and the power rate between LF and HF bands ($\varphi$). The measure $\mathcal{O}$ was considered as the proportion of the spectral power within the DF and its harmonics (considering ±0.5 Hz bandwidth) to the frequency band of interest total spectral power [43]. In this case, as equation (5) shows, only the two first harmonics of the DF were taken into account. For a more global overview of the harmonic structure of the DF, it is introduced a novel parameter $\varphi$ defined as the power ratio between the HF band, which consider the harmonic content, and the LF band, which holds the fundamental frequency, such as equation (6) shows. Therefore, this index could be considered as an estimation of the total harmonic distortion of the $f$-waves signal.

$$\mathcal{O} = \frac{\sum_{i=1}^{3} \sum_{if_0-0.5\text{ Hz}}^{if_0+0.5\text{ Hz}} \mathcal{W}(f)}{\sum_{f=3\text{ Hz}}^{25\text{ Hz}} \mathcal{W}(f)}, \quad i = 1, 2, 3 \tag{5}$$

$$\varphi = \frac{\sum_{f=3f_0/2\text{ Hz}}^{25\text{ Hz}} \mathcal{W}(f)}{\sum_{f=3\text{ Hz}}^{3f_0/2\text{ Hz}} \mathcal{W}(f)} = \frac{\sum \mathcal{W}(HF)}{\sum \mathcal{W}(LF)} \tag{6}$$

The relative power rate of the LF and HF bands to the TF band were also calculated as equations (7) and (8) show, and named as $\varphi'_{LF}$ and $\varphi'_{HF}$, respectively. These metrics provide additional information about the individual spectral weight of the DF band and the harmonic content to the total spectral power of the $f$-waves.

$$\varphi'_{LF} = \frac{\sum_{f=3\text{ Hz}}^{3f_0/2\text{ Hz}} \mathcal{W}(f)}{\sum_{f=3\text{ Hz}}^{25\text{ Hz}} \mathcal{W}(f)} = \frac{\sum \mathcal{W}(LF)}{\sum \mathcal{W}(TF)} \tag{7}$$

$$\varphi'_{HF} = \frac{\sum_{f=3f_0/2\ \text{Hz}}^{25\ \text{Hz}} \mathcal{W}(f)}{\sum_{f=3\ \text{Hz}}^{25\ \text{Hz}} \mathcal{W}(f)} = \frac{\sum \mathcal{W}(HF)}{\sum \mathcal{W}(TF)} \quad (8)$$

Finally, other two parameters novelly proposed in this work to characterize the $f$-waves were the amplitude spectrum area ($AMSA$) [48] and the power spectrum area ($PSA$) [47]. $AMSA$ was estimated by weighting the frequency spectrum amplitudes through the cumulative sum of the product of each individual frequency and its corresponding amplitude, such as equation (9) shows. On the contrary, $PSA$ was computed in the same way as $AMSA$, but using the PSD of the signal instead of the amplitude of the frequency spectrum, see equation (10). These features were calculated over the three frequency bands previously described, so that $f_l$ and $f_u$ refer to the lower and upper frequency limits of each one in equations (9) and (10). Precisely, the evaluation of both parameters in the LF band provided the metrics $AMSA_{LF}$ and $PSA_{LF}$, with $f_l$ and $f_u$ being 3 and $3f_0/2$ Hz, respectively. In the same way, $AMSA_{HF}$ and $PSA_{HF}$ were obtained for the HF band, with $f_l$ and $f_u$ being $3f_0/2$ and 25 Hz, and $AMSA_{TF}$ and $PSA_{TF}$ for the TF band, with boundary frequencies being 3 and 25 Hz, respectively.

$$AMSA = \sum_{f=f_l}^{f_u} \sqrt{\mathcal{W}(f)} f \quad (9)$$

$$PSA = \sum_{f=f_l}^{f_u} \mathcal{W}(f) f \quad (10)$$

#### 2.5.2. Time parameters computed from the $f$-waves

The most common time-domain parameter addressed in other studies is the $f$-waves amplitude [25]. This reference characteristic was estimated in the present work through different methods, according to the indications found in the literature. For instance, it was firstly obtained by computing the root mean square value of the $f$-waves ($\mathcal{A}_{rms}$) [49]. Considering $x(n)$ as the temporal series of a 6 s-length interval of $f$-waves and $N$ as the total number of samples of $x(n)$, $\mathcal{A}_{rms}$ was computed as in the equation (11).

$$\mathcal{A}_{rms} = \sqrt{\frac{1}{N} \sum_{n=1}^{N} |x(n)|^2} \quad (11)$$

Other way to estimate the amplitude of the $f$-waves involved calculating the mean value of the upper and lower envelopes difference, ($\mathcal{A}_{env}$). Briefly, these envelopes of $f$-waves were firstly obtained by identifying maximum peaks of $x(n)$, considering those that are 80% higher than the mean value of the signal and omitting those small peaks that occur in the neighborhood of a larger peak, that is, with a minimum distance limit of the 70% of the AF cycle length (AFCL) between maximum peaks. Then, the upper envelope ($e_{max}(n)$) was estimated applying the modified Akima cubic Hermite interpolation and the lower envelope ($e_{min}(n)$) was computed similarly with the inverse of the signal [50]. Finally, as specified in equation (12), the mean value of the envelopes difference over the length of the $f$-wave interval provided the metric $\mathcal{A}_{env}$.

$$\mathcal{A}_{env} = \frac{1}{N} \sum_{n=1}^{N} |e_{max}(n) - e_{min}(n)| \quad (12)$$

The last algorithm employed for computing the $f$-waves amplitude is described by equation (13). This was based on the average of its maximum peak to peak amplitude obtained from $L$ subsegments of the analyzed signal ($\mathcal{A}_{avg}$) [51]. Since the subsegments should have a long enough duration to include at least one AFCL of the 6 s-length interval of the $f$-waves, 20 subsegments ($L = 20$) of 300 ms were considered for the computation of this variable, so that

$$\mathcal{A}_{avg} = \frac{1}{L} \sum_{w=1}^{L} \left( max\{x_w(n)\} - min\{x_w(n)\} \right), \quad (13)$$

where $\{x_w(n)\}_{w=1,\ldots,L} = \{x(n)\}_{n=(w-1)\frac{N}{L}+1,\ldots,w\frac{N}{L}}$

To mitigate potential influences on ECG amplitude, such as variations in recording gain factors, skin conductivity, electrodes impedance, etc., these metrics are frequently normalized as a percentage of the magnitude of the R-peak [52]. Therefore, the normalized values of $\mathcal{A}_{rms}$, $\mathcal{A}_{env}$, and $\mathcal{A}_{avg}$ were also computed and labeled as $n\mathcal{A}_{rms}$, $n\mathcal{A}_{env}$, and $n\mathcal{A}_{avg}$, respectively.

### 2.6. Statistical analysis and classification performance

The values obtained from the variables described above were represented in mean ± standard deviation format for the patients that remained in SR and those who experienced AF recurrence during the 9-month follow-up of period. The Levene's test verified the homoscedasticity property of each pair of population samples provided by every single metric, and the Lilliefors normality test showed that not all of them followed a normal distribution. A parametric Student's $t$-test was then employed to assess statistical differences between the two patient groups under conditions meeting the assumptions of the test. In instances of non-normal data distributions, a non-parametric Mann–Whitney U-test was utilized instead. Moreover, statistical differences in categorical variables, which were reported as number and population percentage, were evaluated using a Fisher exact test. In all cases, a rejection of the null hypothesis was asserted when the test yielded a significant statistical difference of $p < 0.05$.

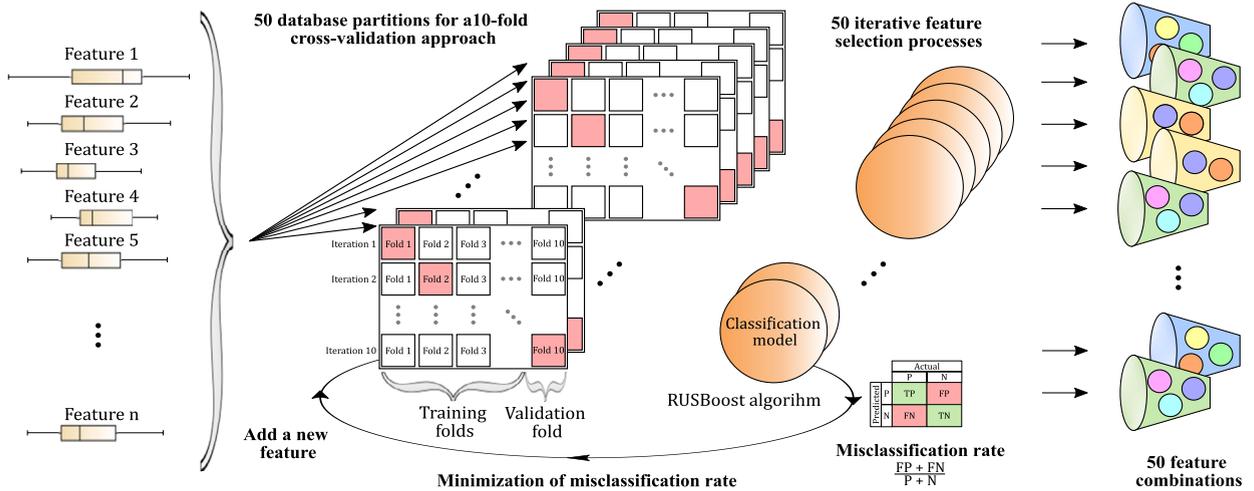

**Fig. 3.** Diagram of the sequential feature selection algorithm.

A 10-fold cross-validation approach was used to assess discriminant ability of each individual parameter [53]. This technique divides the values obtained for all the patients into 10 folds of equal size and repeats a training-validation process 10 times, where a new fold is selected in each iteration to validate a model that was trained with the remaining ones [53]. To ensure the representativeness of each fold for the entire dataset, a stratified division was employed. Moreover, the predictive model trained in every iteration was a data sampling/boosting algorithm. This is an ensemble model based on random undersampling boosting (RUSBoost) of 30 decision trees, each with a limit of 20 splits. It should be noted that this classifier is highly recommended to work with class imbalanced databases [54], such as the one under examination in the current study, where there is a significantly greater proportion of patients who sustained SR after the follow-up period compared to those who experienced a relapse into AF. In fact, the RUSBoost method combines the boosting technique, which is devised to enhance the performance of a weak learner algorithm through an ensemble of classification models, and random undersampling technique, which aims to alleviate the problem of class imbalance. Moreover, unlike oversampling methods, random undersampling approach avoids overfitting, and provides a faster and simpler alternative without losing information when it is combined with a boosting algorithm [54].

In order to mitigate the bias introduced by a singular partition of the data into 10 folds, the validation process for each individual variable was iterated 100 times, with data being reshuffled in every cycle [53]. The classification outcomes derived from each 10-fold cross-validation procedure were summarized through the receiver operating characteristic (ROC) curve. It graphically represents the true positive rate, also known as sensitivity (Se), against the false positive rate (i.e., 1-specificity), across different thresholds applied to the prediction model scores. Whereas Se was established as the percentage of subjects relapsing to AF correctly classified, specificity (Sp) corresponded to the rate of patients maintaining SR who were correctly identified. The threshold selected for optimally separating both patient groups aimed to achieve the best balance between Sp and Se. However, this may not necessarily lead to the highest accuracy (Acc) percentage in correctly classifying patients [55]. The area under the ROC curve (AUC) was also computed as a global overview of the classification performance regardless of that threshold [55]. Additionally, to estimate the rates of true positives and true negatives among its respectively predicted values, the common metrics of positive predictive value (PPV) and negative predictive value (NPV) were also calculated. Ultimately, the mean values of these metrics were computed over the course of the 100 cross-validation iterations conducted.

To explore the complementarity between single variables and improve the prediction of CA outcome, a multivariate study was also conducted. Thus, different subsets of variables were combined through a RUSBoost classifier to obtain several ML-based prediction models. Initially, two models with only the frequency variables (20) and only the time features (6) were built. Next, to exploit all information included in the variables, an additional model containing the 26 analyzed features was generated. Finally, to ensure high predictive ability, reduce complexity in terms of number of variables, and improve clinical interpretability of the results, a wrapper-type feature selection algorithm was used. Precisely, as Fig. 3 depicts, a forward sequential feature selection technique was employed to gradually incorporate variables to a candidate subset that starts empty, with the process continuing until the cross-validated misclassification rate (using 10 folds) was minimized [56]. 50 cycles were conducted to mitigate bias in single data partition and obtain the most characteristic subset of metrics. This involved generating several prediction models with the features more frequently selected. All the models were validated as the individual features, i.e., by repeating 100 processes of a 10-fold cross-validation approach and averaging the resulting performance metrics (Se, Sp, Acc, AUC, PPV, and NPV).

Finally, an asymptotic McNemar's test was used to assess if classification improvements achieved by some prediction models regarding others and regarding the single features were statistically significant. That test is based on comparing the predicted labels obtained by two classification models against the true labels and assessing the statistical significance of the difference in their misclassification rates [57].

**Table 1**
Statistical analysis results of the population baseline demographic and clinical characteristics.

| Clinical Feature | Post-Follow-Up Heart Rhythm | | *p*-Value |
| --- | --- | --- | --- |
| | SR | AF | |
| Number of patients (%) | 103 (68.21%) | 48 (31.79%) | – |
| Male (%) | 79 (76.70%) | 37 (77.08%) | 1.000 |
| Age (years) | 59.37 ± 12.24 | 57.23 ± 12.82 | 0.326 |
| With AF < 1 year (%) | 6 (5.83%) | 6 (12.50%) | 0.198 |
| With AF 1–3 years (%) | 70 (67.96%) | 31 (64.58%) | 0.713 |
| With AF > 3 years (%) | 27 (26.21%) | 11 (22.92%) | 0.841 |
| Body mass index (kg/m$^2$) | 27.79 ± 3.55 | 29.06 ± 4.86 | 0.073 |
| Left atrium diameter (mm) | 44.11 ± 5.70 | 45.65 ± 5.32 | 0.117 |

**Table 2**
Statistical analysis results for the parameters computed from the extracted $f$-waves.

| Feature | Post-Follow-Up Heart Rhythm | | *p*-Value |
| --- | --- | --- | --- |
| | SR | AF | |
| $f_0$ (Hz) | 5.69 ± 1.12 | 6.14 ± 0.99 | **0.009** |
| $\mathcal{W}(f_0)$ (mV$^2$) | 0.00051 ± 0.00092 | 0.00061 ± 0.00092 | 0.058 |
| $f_1$ (Hz) | 11.34 ± 2.25 | 12.28 ± 1.98 | **0.008** |
| $\mathcal{W}(f_1)$ (mV$^2$) | 0.00013 ± 0.00042 | 0.00005 ± 0.00009 | 0.754 |
| $\gamma$ | 2.20 ± 0.77 | 2.80 ± 0.57 | **< 0.001** |
| $b_0$ (Hz) | 0.59 ± 0.14 | 0.62 ± 0.12 | 0.056 |
| $\mathcal{W}(b_0)$ (mV$^2$) | 0.0021 ± 0.0035 | 0.0027 ± 0.0037 | **0.027** |
| $f_m$ (Hz) | 6.20 ± 1.06 | 6.30 ± 0.95 | 0.463 |
| $f_c$ (Hz) | 7.48 ± 0.97 | 7.27 ± 0.74 | 0.111 |
| $\mathcal{W}(f_c)$ (mV$^2$) | 0.00023 ± 0.00045 | 0.00028 ± 0.00044 | **0.040** |
| $\mathcal{O}$ | 0.52 ± 0.15 | 0.53 ± 0.12 | 0.488 |
| $\varphi$ | 0.44 ± 0.30 | 0.24 ± 0.15 | **< 0.001** |
| $\varphi'_{LF}$ | 0.73 ± 0.11 | 0.82 ± 0.07 | **< 0.001** |
| $\varphi'_{HF}$ | 0.27 ± 0.11 | 0.18 ± 0.07 | **< 0.001** |
| $AMSA_{LF}$ (mV · Hz) | 1.64 ± 1.06 | 2.15 ± 1.10 | **0.002** |
| $AMSA_{HF}$ (mV · Hz) | 4.35 ± 3.05 | 4.20 ± 2.67 | 0.919 |
| $AMSA_{TF}$ (mV · Hz) | 5.98 ± 3.91 | 6.34 ± 3.54 | 0.251 |
| $PSA_{LF}$ (mV$^2$ · Hz) | 0.019 ± 0.027 | 0.027 ± 0.031 | **0.007** |
| $PSA_{HF}$ (mV$^2$ · Hz) | 0.017 ± 0.033 | 0.014 ± 0.019 | 0.951 |
| $PSA_{TF}$ (mV$^2$ · Hz) | 0.037 ± 0.057 | 0.040 ± 0.048 | 0.091 |
| $\mathcal{A}_{rms}$ (mV) | 0.033 ± 0.023 | 0.036 ± 0.021 | 0.107 |
| $n\mathcal{A}_{rms}$ | 0.082 ± 0.070 | 0.058 ± 0.034 | 0.064 |
| $\mathcal{A}_{env}$ (mV) | 0.096 ± 0.071 | 0.106 ± 0.068 | 0.147 |
| $n\mathcal{A}_{env}$ | 0.24 ± 0.20 | 0.17 ± 0.10 | 0.072 |
| $\mathcal{A}_{avg}$ (mV) | 0.119 ± 0.085 | 0.128 ± 0.076 | 0.179 |
| $n\mathcal{A}_{avg}$ | 0.30 ± 0.25 | 0.21 ± 0.12 | **0.034** |

## 3. Results

### 3.1. Performance of clinical indices as predictors

The baseline demographic and clinical information for the two considered patient groups, i.e., individuals who sustained SR and those who experienced a relapse to AF during the follow-up, is represented in Table 1. Of note is that none of the analyzed indices provided statistically significant differences.

### 3.2. Univariate analysis

Average and standard deviation values of the features computed to characterize the $f$-waves in the frequency- and time-domains are presented in Table 2 for the two groups of patients. This table also shows the *p*-value provided in the evaluation of the statistical differentiation ability of each parameter to separate the two population samples. As can be seen, only some features obtained a value lower enough to reject the null hypothesis that states equality of mean of both groups. Regarding the spectral features addressed in previous works, $f_0$, $f_1$, $\gamma$, and $\mathcal{W}(b_0)$ yielded statistically significant differences between the two patient groups, with a lower mean value for those maintaining SR. Among the novel spectral parameters introduced in the study, $\mathcal{W}(f_c)$, the three power rate metrics ($\varphi$, $\varphi'_{LF}$, and $\varphi'_{HF}$), and AMSA and PSA for the LF band were able to statistically differentiate both groups of patients. The variables $\mathcal{W}(f_c)$, $AMSA_{LF}$, and $PSA_{LF}$ provided a lower mean result for the patients who sustained SR than for those presenting AF recurrence within the first 9 months. A similar result was also noticed for the index $\varphi'_{LF}$, whereas $\varphi'_{HF}$ reported the contrary trend, thus also leading to larger values of $\varphi$ for the patients who preserved SR than for those who experienced AF recurrence.

**Table 3**
Univariate classification results.

| Feature | Se (%) | Sp (%) | Acc (%) | AUC (%) | PPV (%) | NPV (%) |
|---|---|---|---|---|---|---|
| $f_0$ | 61.23 | 61.15 | 61.17 | 63.69 | 42.34 | 77.19 |
| $\gamma$ | 61.46 | 61.58 | 61.54 | 61.19 | 42.71 | 77.42 |
| $\varphi$ | 63.77 | 63.86 | 63.83 | 66.06 | 45.13 | 79.09 |
| $\varphi'_{LF}$ | 63.17 | 63.25 | 63.23 | 67.95 | 44.48 | 78.66 |
| $\varphi'_{HF}$ | 62.79 | 62.79 | 62.79 | 67.93 | 44.02 | 78.36 |
| $AMSA_{LF}$ | 57.83 | 57.90 | 57.88 | 56.89 | 39.03 | 74.66 |
| $n\mathcal{A}_{rms}$ | 54.92 | 54.91 | 54.91 | 56.89 | 36.21 | 72.33 |
| $n\mathcal{A}_{env}$ | 52.18 | 52.22 | 52.22 | 53.86 | 33.73 | 70.09 |
| $n\mathcal{A}_{avg}$ | 50.64 | 50.56 | 50.60 | 53.18 | 32.31 | 68.73 |

**Table 4**
Multivariate classification results.

| Features in the Model | Se (%) | Sp (%) | Acc (%) | AUC (%) | PPV (%) | NPV (%) |
|---|---|---|---|---|---|---|
| All frequency features (20) | 68.42 | 68.41 | 68.41 | 72.32 | 50.23 | 82.30 |
| All time features (6) | 55.44 | 55.58 | 55.54 | 59.28 | 36.77 | 72.80 |
| All features (26) | 70.40 | 70.40 | 70.40 | 75.04 | 52.57 | 83.61 |
| $\varphi, AMSA_{LF}$ | 73.77 | 73.73 | 73.74 | 76.04 | 56.68 | 85.78 |
| $\varphi, AMSA_{LF}, n\mathcal{A}_{avg}$ | 78.52 | 78.55 | 78.54 | 79.82 | 63.05 | 88.70 |
| $\varphi, AMSA_{LF}, n\mathcal{A}_{avg}, n\mathcal{A}_{rms}$ | 77.65 | 77.67 | 77.66 | 78.49 | 61.84 | 88.17 |
| $\varphi, AMSA_{LF}, n\mathcal{A}_{avg}, n\mathcal{A}_{env}$ | 77.31 | 77.31 | 77.31 | 78.58 | 61.36 | 87.97 |
| $\varphi, AMSA_{LF}, n\mathcal{A}_{avg}, f_0, \varphi'_{LF}$ | 76.67 | 76.65 | 76.66 | 81.60 | 60.48 | 87.58 |
| $\varphi, AMSA_{LF}, n\mathcal{A}_{avg}, f_0$ | 76.44 | 76.44 | 76.44 | 81.07 | 60.19 | 87.44 |
| $\varphi, AMSA_{LF}, n\mathcal{A}_{avg}, \varphi'_{LF}$ | 76.44 | 76.41 | 76.42 | 79.81 | 60.16 | 87.43 |

On the other hand, none of the non-normalized $f$-waves amplitude measures yielded noteworthy statistical distinctions between the two patient groups. Contrarily, the normalized versions of these metrics reported values of statistical significance close to be significant, such as $n\mathcal{A}_{rms}$ and $n\mathcal{A}_{env}$, or directly lower than 0.05, such as $n\mathcal{A}_{avg}$. In all the cases, they provided higher mean values of $f$-waves amplitude for those patients who maintained SR.

To serve as a reference for comparison with the results obtained in multivariate analysis, the classification performance of the single-feature models based on RUSBoost algorithm of those metrics that play a relevant role in the generated multivariate models is presented in Table 3. The well-known $f_0$ and $\gamma$, and the newly proposed $\varphi$, $\varphi'_{LF}$, $\varphi'_{HF}$, and $AMSA_{LF}$ presented the best individual classification results, being the novel features superior to the preceding ones. In particular, the index $\varphi$ presented the highest values of Acc, Se, Sp, and AUC, which were near or higher than 64%. Compared with these frequency-domain parameters, those estimating $f$-waves amplitude reported a poorer classification performance with metrics of Acc, Se, Sp, and AUC between 50 and 57%.

*3.3. Multivariate analysis*

The performance classification for all the generated RUSBoost-based multivariate prediction models is presented in Table 4. As can be seen, those based on only time features reported the poorest outcomes, with values of Acc and AUC of about 55% and 60%, respectively. Compared with this model, the two based on only frequency variables and on all 26 features provided statistically better performances (according to a McNemar's test), with Acc and AUC values around 68–70% and 72–75%, respectively. Finally, the selection of optimal features also allowed us to generate prediction models with statistically better performance regarding that combining all the 26 features. To this respect, after computing the sequential feature selection algorithm for 50 different database partitions, the most predictive parameters to be combined through a RUSBoost classifier were $\varphi$, $AMSA_{LF}$, and $n\mathcal{A}_{avg}$. The resulting model provided the best classification result, with values of Acc, and AUC of about 78.5–80%. The improvement reached regarding the three single features was quite notable and statistically significant. By way of illustration, Fig. 4 compares the performance of the prediction model with the three single features through a radar chart, where its vertices represent the classification performance metrics. Of note is that the prediction model yielded PPV, NPV, AUC, Se, Sp, and Acc values between 10% and 20% greater than the single variables.

In addition to this last model, others were also evaluated by incorporating some single parameters sometimes chosen in the conducted procedure of automated feature selection. Thus, the subset of the indices $\varphi$, $AMSA_{LF}$, and $n\mathcal{A}_{avg}$ was complemented by the variables $n\mathcal{A}_{rms}$, $n\mathcal{A}_{env}$, $f_0$, and $\varphi'_{LF}$, such as Table 4 displays. The obtained models did not overcome the performance of the previous one, and only provided slightly and not statistically significant, lower classification outcomes. Nonetheless, they presented a performance statistically better than the corresponding single variables and the model combining all the 26 parameters.

**4. Discussion**

CA for AF patients is a well-established treatment, especially suitable for reducing arrhythmia-related symptoms, as it is aimed at terminating AF episodes and maintaining SR [2]. However, the mid-term effectiveness of PVI in individuals suffering from persistent

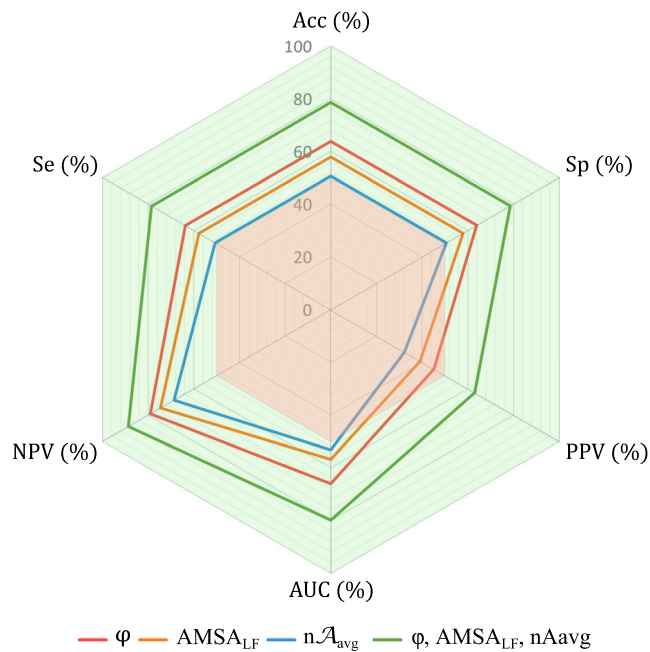

**Fig. 4.** Performance comparison on a radar chart between the best multivariate model and the univariate models of the individual features that it combines.

AF is limited, despite applying additional isolation lines in the atrial substrate to eliminate and isolate arrhythmogenic regions or performing repeated procedures [58]. This situation places a high burden on the health systems becoming one of the most significant cause of rising healthcare costs in Europe [59]. In fact, the presence of cardiac disease concomitant with the current major medical challenge caused by the COVID-19 virus lead to worse outcomes associated with more complications and further aggravate the burden on the healthcare systems [60,61]. Nonetheless, CA remains a widely used therapy for persistent AF patients. The preoperative anticipation of its mid-term success before the procedure is of increasing clinical interest [22], as it may prevent unnecessary risks for AF patients and improve the management of healthcare costs, among other benefits [17].

The study of certain clinical characteristics of AF patients has been repeatedly interpreted as a source of information that may be useful beyond the diagnosis of the disease. To date, some of the clinical indices of AF recurrence after CA that have been examined are based on the patient's medical history (e.g., age, AF duration, and concomitant diseases) [18], on anatomical and functional parameters of the atria (e.g., left atrial diameter, atrial wall thickness, and left atrial volume) [62], on the measurement of the level of plasma substances secreted by the heart to regulate the circulatory system [63], or on inflammatory markers associated with atrial remodeling [64]. However, the exhaustive review of these clinical indices conducted by Balk et al. [18] concluded the existence of great controversy on their predictive capacity of CA outcome. In this line, none of the demographic and clinical indices assessed in the present study was able to statistically differentiate between patients maintaining SR and individuals who experienced AF recurrence within the first 9 postoperative months.

With the same purpose of anticipating CA outcome, other authors were pioneers in analyzing the electrical activity of the atria. In particular, the DF, or its inverse (the AFCL), and the amplitude of the $f$-waves have reported a promising ability to predict the CA success [24,25]. On the one hand, Haïssaguerre et al. [65] evaluated the variation of the AFCL during the CA procedure and were able to verify that its increase through the intervention was much more significant in those patients for whom CA terminated the arrhythmia. Takahashi et al. [66] observed an association between the termination of AF with a decrease in the DF during the CA procedure, and Yoshida et al. [67] found that such decrease was also correlated with the effectiveness of the intervention in subjects suffering from persistent AF. Yoshida et al. [68] continued studying the evolution of the DF during the CA procedure and found that a decrease in its value greater than 11% was able to predict postoperative absence of AF recurrence with an accuracy about 74%. Finally, among many other articles, Di Marco et al. [69] also demonstrated the ability of the AFCL as a predictor of mid-term SR restoration after CA of persistent AF ($p < 0.01$). However, despite the great interest of these results, it should be noted that measures of DF and AFCL were developed invasively through electrograms obtained during the intervention, thus involving the impossibility of using these markers as preoperative predictors and preventing patients from unnecessary risks.

On the other hand, Cheng et al. [28] observed that the existence of low $f$-waves amplitude in lead V1 ($< 0.123$ mV) of the preoperative ECG was predictive of a high probability of unsuccessful CA intervention (with values of Se and Sp of about 68% and 64%, respectively). These findings align with the outcomes reported in a prior work, where a greater amplitude of the $f$-waves, measured in lead V1, was predictive of both the termination of the arrhythmia during the CA procedure and the maintenance of SR after the first postoperative year ($p = 0.004$) [38]. This finding has been associated with the fact that the patients with a more disorganized fibrillatory activity were characterized by having multiple activation foci that generated wavefront propagations in different directions in the atria. This situation may cause a greater number of wavefront collisions, so that a lower $f$-wave amplitude was registered in the ECG [49]. However, these studies [28,38] performed manual measurements of the $f$-wave amplitude, which entails a high degree of subjectivity.

To address these constraints, some authors have proposed the automated evaluation of similar frequency and amplitude measures in the surface ECG signal preoperatively acquired before the CA procedure. To this respect, Matsuo et al. [70] corroborated the predictive potential of the non-invasively estimated AFCL, because it yielded a higher preoperative value for patients exhibiting sustained long-term SR after CA ($p < 0.001$, AUC = 0.88). In the same line, the DF mainly estimated from lead V1 also provided a promising ability to anticipate CA result, with lower values suggesting lower probability of relapsing to AF and lesser degree of electrical remodeling in the atria [43,44]. As can be seen in Table 2, a similar trend was noticed in the results of $f_0$ (and therefore of $f_1$) obtained in the present work. Although other previous studies analyzing the DF did not find statistically significant results, it should be noted that they enrolled notably smaller databases of persistent AF patients (about 20 subjects) [26,71].

Automated estimation of $f$-wave amplitude to anticipate CA success has previously been addressed by some authors [50,72]. Unlike the DF, in this case more controversial results are found. Whereas some works did not found statistical differences between subjects maintaining SR and relapsing to AF within the postoperative follow-up when single leads were analyzed [50], others have yielded classification results comparable to some clinical indices [72]. In the present study, the normalized estimates of the $f$-waves amplitude computed from different algorithm, i.e., $n\mathcal{A}_{rms}$, $n\mathcal{A}_{env}$, and $n\mathcal{A}_{avg}$, provided the same tendency than in the studies where manual measurements were analyzed [38,28]. However, compared with these previous works, the indices $n\mathcal{A}_{rms}$, $n\mathcal{A}_{env}$, and $n\mathcal{A}_{avg}$ reported a considerably more limited predictive capacity, around 55% (Table 3).

Some recent works have extended the spectral characterization of the non-invasive $f$-waves by considering additional parameters, such as $f_1$, $\mathcal{W}(f_0)$, $\mathcal{W}(f_1)$, $b_0$, $\mathcal{W}(b_0)$, $\mathcal{O}$, $f_m$, and $\gamma$, among others [26,73]. As in the present work, whereas most of these parameters did not exhibit differences between the patient groups that attained statistical significance, the index $\gamma$ proved a high discriminant ability between both groups (see Table 3). It returned lower values for the patients maintaining SR, thus highlighting the existence of a smaller gap between the spectral power of the DF and its first harmonic. Several authors have previously associated this result with the existence of stronger DF harmonics, thus suggesting more organized $f$-waves, more organized atrial conduction, and consequently lower likelihood of AF recurrence [26,66,74]. However, this parameter ignores the information presented by the DF harmonic components beyond the first one, and novel indices for a more global estimation of the harmonic structure of the $f$-waves were proposed, such as $\varphi$, $\varphi'_{LF}$, and $\varphi'_{HF}$.

The metric $\varphi$ quantifies overall importance of the harmonic structure in the HF band in comparison with the spectral content in the LF band around the DF. It showed the best performance among all the analyzed single parameters, providing higher values for those patients who benefited from the CA procedure and maintained SR along the first postoperative months. This finding aligns with the results reported by the new indices $\varphi'_{LF}$ and $\varphi'_{HF}$, which assessed the PSD content of both the LF and HF bands in relation to the TF band, respectively. Thus, $\varphi'_{LF}$ provided higher values for those patients who experienced AF recurrence than for those maintaining SR. This higher relative PSD concentration in the LF band regarding the TF band also agree with the absolute higher values of the PSD in $b_0$ provided by the $\mathcal{W}(b_0)$ feature and suggests a stronger component of the DF in the case of patients more likely to relapse to AF. Contrarily, higher values of $\varphi'_{HF}$ were obtained for the group of patients who maintained SR. This finding is in agreement with other studies suggesting that the presence of stronger harmonics in the $f$-waves is associated with the termination of the arrhythmia during CA [66], as well as with the mid-term maintenance of SR [26,73,75].

Other novel features analyzed in the present study to anticipate the success of CA were $f_c$, $\mathcal{W}(f_c)$, $AMSA$, and, $PSA$. They have been previously used for defibrillation outcome prediction in patients with ventricular fibrillation, reporting auspicious results [48,51]. However, despite certain morphological similarity noticed between the $f$-waves and the ECG signal in ventricular fibrillation, the classification results exhibited by these parameters were lower than expected. Only $AMSA_{LF}$ and $PSA_{LF}$ reported results yielding differences between the two patient groups that were statistically significant, but with limited values of Acc and AUC of about 58% or smaller. Nonetheless, $AMSA_{LF}$ proved a key role in the generated prediction models, because it complemented the parameter $\varphi$ to obtain a statistically significant improvement in classification of about 15% in terms of all performance metrics (see Table 4). Moreover, when the amplitude variable $n\mathcal{A}_{avg}$ was added to the subset consisting of $\varphi$ and $AMSA_{LF}$, the model still increased classification up to the best values of Acc and AUC of 78–80%. Bearing in mind that $AMSA_{LF}$ integrates time and frequency information, and $\varphi$ and $n\mathcal{A}_{avg}$ respectively capture isolated frequency and time $f$-waves characteristics, such result suggests that the combination of features from both domains through a ML-based classifier is helpful to improve CA outcome. Accordingly, the prediction model combining the three features reported a statistically superior classification performance than the models combining only frequency or time variables, with improvements larger than 20% and 30% in all performance metrics, respectively.

Additionally, although caution is advised in direct comparison of classification results from different experimental setups, it is worth noting that the prediction model based on the indices $\varphi$, $AMSA_{LF}$, and $n\mathcal{A}_{avg}$ also provided results comparable or even larger than in previous works dealing with ECG-based prediction of CA outcome. To this respect, values of Acc lower than 70% have been reported by previous studies conducted on databases with a similar number of patients than the present work (i.e., 120 or more) [43,44,72,73]. Moreover, although other studies have reported comparable or slightly larger values of Acc of about 75–85% [26–28,76–78], they analyzed highly limited datasets of less than 65 patients with persistent AF, and did not use a robust validation approach. Precisely, most of them were based on a resubstitution validation approach, utilizing the same dataset of patients for both training and testing, thus resulting in poorly general and notably inflated classification results [53]. Furthermore, an interesting characteristic of the proposed model, which was not always presented by previous ones [27,78], is that only combined three single features, thus preserving clinical interpretability.

## 5. Limitations

The present study has some drawbacks that need to be taken into account. Thus, although the database of the study is broad enough to be scalable to persistent AF patient population, the subset of patients maintaining SR is higher than the proportion of those who relapsed to AF, leading to an unbalanced database. Nonetheless, this imbalance is common in clinical practice, so that the study represents a real sample of the persistent AF patient's population undergoing CA to treat the arrhythmia. Despite this, a RUSBoost machine learning algorithm used to obtain the classification models was specifically considered for mitigating this class imbalance problem [54]. On the other hand, the characterization of the $f$-waves was restricted to lead V1. Even though this single lead provides the best signal to enhance AF activity regarding its ventricular counterpart [31,38], later studies have demonstrated that the combination of metrics from different ECG leads could improve prediction outcomes [50,71,78,79]. Hence, future improvements in predicting CA outcome for persistent AF patients could involve multi-variable and multi-lead analyses, incorporating the proposed frequency and time-domain features. Finally, in addition to the demographic and clinical variables analyzed in Table 1, others such as the presence of hypertension, diabetes, obstructive sleep apnea, structural heart disease, underlying heart disease, hyperthyroidism, smoking history, and pulmonary disease have been considered by some authors as predictive of CA outcome [22]. Unfortunately, these variables were not available for all the patients enrolled in the study and could not be assessed.

## 6. Conclusions

The combination of frequency- and time-domain features of the $f$-waves through a ML-based classifier, such as RUSBoost, has significantly outperformed the capacity of single predictors and other previously proposed predictive models in anticipating the outcome of CA procedure in subjects suffering from persistent AF. Furthermore, the novel parameter $\varphi$ has been brought to light as potential predictor, corroborating the relevance of the harmonic structure of the $f$-waves in helping to differentiate between patients sustaining SR and presenting AF recurrence within the first post-operative 9 months. Also, the incorporation of $\varphi$ with other frequency- and time-domain features, as the variable $AMSA_{LF}$ and the normalized $f$-waves amplitude $n\mathcal{A}_{avg}$, in a multivariate model has returned the most promising predictive outcome. Therefore, the synthesis of information both from time and frequency characteristics of the $f$-waves could improve clinical assistance in persistent AF treatment to avoid unnecessary risks associated with CA in subjects with low likelihood of successfully responding to the intervention.

**List of acronyms**

| Acronym | Definition |
| --- | --- |
| Acc | Accuracy |
| AMSA | Amplitude spectrum area |
| AUC | Area under the ROC curve |
| AF | Atrial fibrillation |
| AFCL | Atrial fibrillation cycle length |
| CA | Catheter ablation |
| DF | Dominant frequency |
| ECG | Electrocardiogram |
| $f$-waves | Fibrillatory waves |
| HF | High-frequency |
| LF | Low-frequency |
| ML | Machine learning |
| NPV | Negative predictive value |
| PPV | Positive predictive value |
| PSD | Power spectral density |
| PSA | Power spectrum area |
| PVI | Pulmonary vein isolation |
| RUSBoost | Random undersampling boosting |
| ROC | Receiver operating characteristic |
| Se | Sensitivity |
| SR | Sinus rhythm |
| Sp | Specificity |
| TF | Total-frequency |

**Funding**

This research was financially supported from public grants PID2021-00X128525-IV0, PID2021-123804OB-I00, and TED2021-130935B-I00 of the Spanish Government 10.13 039/501100011033 jointly with the European Regional Development Fund (EU), SBPLY/21/180501/000186 from Junta de Comunidades de Castilla-La Mancha, and AICO/2021/286 from Generalitat Valenciana. Furthermore, Pilar Escribano holds a 2020-PREDUCLM-15540 predoctoral scholarship, co-financed by the operating program of the European Social Fund (ESF) 2014–2020 of Castilla-La Mancha.

**Ethics statement**

The study was conducted in accordance with the Declaration of Helsinki guidelines and complied with Spanish regulations, including Law 14/2007, 3rd of July, on Biomedical Research. Ethical approval for the study was granted by the Ethical Review Board of University Hospitals of Toledo and Albacete, Spain, under protocol code 5064, 1 December 2020. Written informed consent was obtained from all the subjects.

**CRediT authorship contribution statement**

**Pilar Escribano:** Writing – original draft, Visualization, Software, Methodology, Investigation, Data curation, Conceptualization. **Juan Ródenas:** Writing – review & editing, Supervision, Software, Data curation, Conceptualization. **Manuel García:** Writing – review & editing, Supervision, Software, Data curation, Conceptualization. **Miguel A. Arias:** Writing – review & editing, Resources. **Víctor M. Hidalgo:** Writing – review & editing, Resources. **Sofía Calero:** Writing – review & editing, Resources. **José J. Rieta:** Writing – review & editing, Validation, Supervision, Methodology, Data curation, Conceptualization. **Raúl Alcaraz:** Writing – review & editing, Validation, Supervision, Methodology, Data curation, Conceptualization.

**Declaration of competing interest**

The authors declare that they have no known competing financial interests or personal relationships that could have appeared to influence the work reported in this paper.

**Data availability**

The data supporting reported results and presented in this study are available on request from the corresponding author.